\documentclass[12pt]{article}
\usepackage{epsfig}
\font\mybb=msbm10 at 12pt
\font\mybbsub=msbm10 at 10pt

\def\bb#1{\hbox{\mybb#1}}
\def\bbsub#1{\hbox{\mybbsub#1}}

\def\ZZ {\bb{Z}}
\def\RR {\bb{R}}
\def\PP {\bb{P}}
\def\ZZsub {\bbsub{Z}}
\def\RRsub {\bbsub{R}}

\def\CC {\bb{C}}
\newcommand{\vs}[1]{\vspace{#1 mm}}

\begin{document}
\begin{titlepage}

\setcounter{page}{0}
\begin{flushright}
\mbox{$\begin{array}{l}
\mbox{KEK Preprint 2000-107}\\
\mbox{hep-th/0009240}\\
\mbox{September 2000}\\
\mbox{H}
\end{array}$
}
\end{flushright}

\vs{5}
\begin{center}
{\Large\bf Noncompact Gepner Models for Type~II Strings on a 
Conifold and an ALE Instanton\footnote{Talk given at 30th 
International Conference on High-Energy Physics (ICHEP 2000), 
Osaka, Japan, 27 Jul - 2 Aug 2000.} 
\\}

\vs{10}

{\large
Shun'ya Mizoguchi\footnote{shunya.mizoguchi@kek.jp}} \\
\vs{5}
{\it Institute of Particle and Nuclear Studies \\
High Energy Accelerator Research Organization (KEK) \\
Oho 1-1, Tsukuba, Ibaraki 305-0801, Japan} \\
\end{center}
\vs{10}
\centerline{{\bf{Abstract}}}
\vskip 3mm
We construct modular invariant
partition functions for type~II strings on a conifold and a singular 
Eguchi-Hanson instanton by means of the $SL(2,\RR)/U(1)$ 
version of Gepner models. In the conifold case, we find an extra 
massless hypermultiplet in the IIB spectrum and argue that it 
may be identified as a soliton. In the Eguchi-Hanson case, 
our formula is new and different from the earlier result, in particular 
does not contain graviton. The lightest IIB fields are combined into  
a six-dimensional $(2,0)$ tensor multiplet with a negative mass square. 
We give an interpretation to it as a doubleton-like mode.


\end{titlepage}

\baselineskip=18pt
\setcounter{footnote}{0}

\section{Introduction}
The recognition of the importance of physics near the singularity 
in the moduli space is one of the highlights in the recent developments 
of string theory. Well-known examples are the conifold singularity 
of Calabi-Yau three-folds and the ADE singularities of K3 surfaces. 
In both cases, type~IIA or IIB strings acquire  
extra massless solitons due to various D-branes wrapped around 
the vanishing cycles of the singularity\cite{S,W}. 

In this contribution we construct modular invariant 
partition functions for the CFTs which describe those theories on 
such singularities.\footnote{
We do not consider any D-brane probe.
}
At first sight, such an attempt might look like nonsense because one 
would expect non-perturbative effects near the singularity, 
where the CFT description of the strings on a {\it compact} space 
breaks down. However, we may still expect some dual perturbative 
CFT description for those theories\cite{BVS}, where the 
non-perturbative effects in one theory are studied perturbatively 
in the other theory. The idea is to ``pinpoint'' those theories by 
using an abstract CFT approach, just as we can use ordinary Gepner 
models to describe some special points of moduli space of ordinary 
Calabi-Yau compactifications.

This paper is organized as follows. 
In section 2, we construct a partition function 
for type~II strings on a conifold\cite{M} and argue that the extra 
massless hypermultiplet that appears in the spectrum may be identified
as a soliton. In section 3, we address an issue in the known 
partition function formulas for the singular ALE spaces. 
In section 4, we present a new modular invariant partition 
function for type~II strings on a singular Eguchi-Hanson instanton, the 
simplest ($A_1$) four-dimensional ALE space in the ADE classification. 
Our formula is new and different from the earlier result\cite{OV,ES}. The 
last section is devoted to some concluding remarks.

\section{Conifold}
Let us begin with four-dimensional type~II theories ``compactified''
on a conifold. The CFT for the four-dimensional part is a free $N=2$ 
SCFT, while the conifold part is the $c=9$, 
$SL(2,\RR)/U(1)$ Kazama-Suzuki model\cite{GV}. The clue  
that leads to the relation between a conifold and the 
$SL(2,\RR)/U(1)$ SCFT may be found in the equation 
of the deformed conifold (See ref.\cite{OV} for further 
explanations and references.): 
\begin{equation}
z_1^2+z_2^2+z_3^2+z_4^2=\mu,
\end{equation}
where $z_i$ $(i=1,\ldots,4)$ are the coordinates of 
$\CC^4$. $\mu\rightarrow 0$ is the conifold limit. 
To apply the abstract CFT approach we replace $\mu$ with $\mu z_5^{-1}$,
where $z_i$ $(i=1,\ldots,5)$ are now thought of as the coordinates 
of $\PP^4$. The negative power of $-1$ has 
been determined by the Calabi-Yau condition, and interpreted as 
the $SL(2,\RR)/U(1)$ $N=2$ model with level 
$k=-(-1-2)=3$. Its central charge is $c=\frac{3k}{k-2}=9$.
Thus, unlike ordinary Gepner models, the necessary central charge 
for the internal CFT is supplied by a single 
$SL(2,\RR)/U(1)$ CFT.

In fact, if $c>3$, {\it any} unitary representation for the $N=2$ 
superconformal algebra can be constructed from the 
$SL(2,\RR)/U(1)$ coset\cite{DLP}. It means that
we are allowed to use any $c=9$ representations. What's the 
criteria for the representations to be chosen? Our guideline is modular 
invariance and spacetime supersymmetry.  

The generic (nondegenerate) $N=2$, $c=9$ superconformal characters 
are given by  
\begin{equation}
\mbox{Tr}q^{L_0}y^{J_0}
=q^{h+1/8}y^Q\frac{\vartheta_3(z|\tau)}{\eta^3(\tau)}
\label{generic_NScharacter}
\end{equation}
(NS sector), where 
$q=\exp(2\pi i \tau)$ and $y=\exp(2\pi i z)$.
To improve the modular behavior of the monomial factor 
$q^{h+1/8}$, we consider a countable set of infinitely many 
representations so that the sum of the monomial factors forms
a certain theta function. (If the $c=9$ CFT is realized as an $N=2$ 
Liouville$\times S^1$ system\cite{N=2Liouville,GK}, this operation 
amounts to the momentum summation along $S^1$.) It is still not enough 
to construct a modular invariant combination because the numbers of 
theta and $\eta$ functions are not the same. (The $\sqrt{\tau}$ 
factors do not cancel in the modular $S$ transformation.) To cure this 
problem, we consider {\it continuously many} representations for each 
$U(1)$ charge in the above theta function and integrate over 
them with the Gaussian weight (Liouville 
momentum integration).
 
Which theta functions should we use? To determine them we require 
spacetime supersymmetry. We now have three theta functions: one from 
the free complex fermion for the two transverse spacetime dimensions, 
one from the Jacobi theta function in the $N=2$ characters, and one 
from the character summation above. They must be GSO projected in 
an appropriate way to give a vanishing partition function for the 
theory to be supersymmetric. Therefore, we need some identities among  
the products of three theta functions. One solution has been known 
for a long time\cite{BG}: 
\begin{eqnarray}
\Lambda_1(\tau)&\equiv&\Theta_{1,1}(\tau,0)
\left(\vartheta_3^2(0|\tau)+\vartheta_4^2(0|\tau)\right)
-\Theta_{0,1}(\tau,0)
\;\vartheta_2^2(0|\tau)~=~0,
\label{theta_id}
\end{eqnarray}
where 
\begin{eqnarray}
\Theta_{m,1}(\tau,z)&=&
\sum_{n\in\ZZsub}q^{(n+m/2)^2}y^{(n+m/2)},
\end{eqnarray}
($m=0,1$)
are the level-1 $SU(2)$ theta functions. 
There is another solution\cite{M}:
\begin{eqnarray}
\Lambda_2(\tau)&\equiv&\Theta_{0,1}(\tau,0)
\left(\vartheta_3^2(0|\tau)-\vartheta_4^2(0|\tau)\right)
-\Theta_{1,1}(\tau,0)
\;\vartheta_2^2(0|\tau)~=~0,
\label{theta_id'}
\end{eqnarray}
which is nothing but the modular $S$ transform of (\ref{theta_id}).
Their modular transformations are 
\begin{equation}
\Lambda_1(\tau+1)=i\Lambda_1(\tau),~~
\Lambda_2(\tau+1)=-\Lambda_2(\tau),
\end{equation}
and
\begin{eqnarray}
\Lambda_1\!\left(-\frac1{\tau}\right)&=&
 \frac{e^{-\frac{3\pi i}4}\tau^{\frac32}}{\sqrt{2}}
\left(-\Lambda_1(\tau)+\Lambda_2(\tau)\right),\nonumber\\
\Lambda_2\!\left(-\frac1{\tau}\right)&=&
 \frac{e^{-\frac{3\pi i}4}\tau^{\frac32}}{\sqrt{2}}
\left(\Lambda_1(\tau)+\Lambda_2(\tau)\right).
\end{eqnarray}
Thus 
$
\left|\Lambda_1(\tau)/\eta^3(\tau)\right|^2
+\left|\Lambda_2(\tau)/\eta^3(\tau)\right|^2
$
is modular invariant. Including the other contributions as well, 
we obtain the following modular invariant partition function: 
\begin{eqnarray}
\phantom{}Z_{\mbox{\scriptsize conifold}}
=\!\!\int\!\frac{d^2\tau}{(\mbox{Im}\tau)^2}
	\frac1{(\mbox{Im}\tau)^{\frac32}\left|\eta(\tau)\right|^6}
\left[\left|\Lambda_1(\tau)/\eta^3(\tau)\right|^2
\!+\left|\Lambda_2(\tau)/\eta^3(\tau)\right|^2\right]\!.
\end{eqnarray}
One may easily read off from $Z_{\mbox{\scriptsize conifold}}$ 
which $N=2$, $c=9$ representations are used (Figure 1; see ref.\cite{M}
for the R-sector.).
\begin{figure}
\begin{center}
\epsfig{file=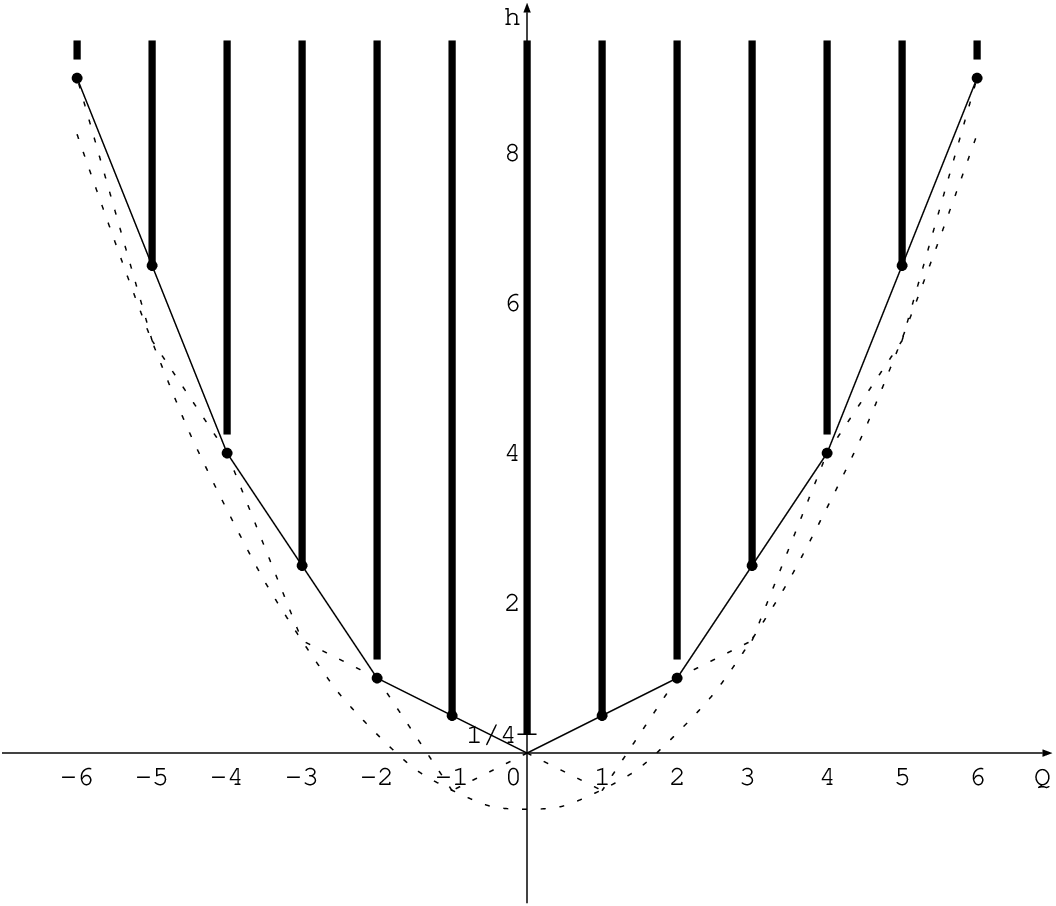, width=100mm}%
\end{center}
\caption{$N=2$, $c=9$ representations used as the internal 
CFT (NS-sector).}
\end{figure}

The following observations support that 
$Z_{\mbox{\scriptsize conifold}}$ is really the 
partition function for type~II strings on a conifold:
\vskip 1ex\noindent
\raisebox{-1mm}{\Huge $\cdot$} First of all, the spectrum exhibits 
a continuum. This is 
due to the integration over the $N=2$ representations 
(Liouville momentum integration), which is a consequence of 
the requirement from modular invariance. Therefore, spacetime, 
which was initially supposed to be four-dimensional, becomes 
five-dimensional effectively. Another related observation is that 
the graviton, the dilaton and the $B$-field correspond to the 
$(h,Q)=(1/4,0)$ point, and hence are massive.  This is a generic $N=2$ 
representation and corresponds to a non-normalizable (``principal 
unitary series'') 
$SL(2,\RR)$ representation. This is certainly 
different from the ordinary CFT description of the Calabi-Yau 
``{\it compact}''-ifications.
\vskip 1ex\noindent
\raisebox{-1mm}{\Huge $\cdot$} 
Consider type~IIB theory. 
We have one chiral- and one 
anti-chiral primary fields of $h=1/2$ in the 
NS-sector (Figure 1). They (together with their spectral flows) 
give rise to a massless $N=2$, $U(1)$ vector multiplet and a 
hypermultiplet.
The vector multiplet agrees with the Hodge number $h_{2,1}=1$ 
of the deformed conifold, while the hypermultiplet is an extra 
one.\footnote{
This is neither the universal hypermultiplet, nor 
the one coming from the K\"{a}hler form on the conifold; the 
dilaton is paired with the $(h,Q)=(1/4,0)$ representation, while 
the K\"{a}hler form does not have a compact support and hence 
will correspond to a continuous representation of 
$SL(2,\RRsub)$.
}
We will argue that this hypermultiplet may be identified as the 
famous massless soliton\cite{S} coming from the wrapped D3 brane. 
The extra massless fields are due to the second chiral primary 
field $(h,Q)=(1,2)$, which exhibit a gap above.  
Technically, the representations on the boundary of the unitarity 
region are degenerate representations,
and hence are smaller than the generic ones.  
The irreducible character for $(h,Q)=(1/2,1)$ is given by 
\begin{equation}
\mbox{Tr}q^{L_0}y^{J_0}|_{(h,Q)=(\frac12,1)}
=\frac{q^{\frac12+\frac18}y}{1+q^{\frac12}y}
\frac{\vartheta_3(z|\tau)}{\eta^3(\tau)}
\label{degenerate_NScharacter}
\end{equation}
and not in the generic form like (\ref{generic_NScharacter}). 
In fact, their difference is ($(h,Q)=(\frac12,1)$)
\begin{eqnarray}
(\ref{generic_NScharacter})
-(\ref{degenerate_NScharacter})
=\frac{q^{1+\frac18}y^2}{1+q^{\frac12}y}
\frac{\vartheta_3(z|\tau)}{\eta^3(\tau)},
\end{eqnarray}
which is precisely the irreducible character for 
$(h,Q)=(1,2)$\cite{ET}.\footnote{
Degenerate representations on the boundary of the unitarity region 
correspond to discrete series of $SL(2,\RRsub)$\cite{DLP}.
}
(And similarly for the anti-chiral primary fields; they are shown 
by the dots in Figure 1.) Therefore, a generic representation with 
$Q=1$ splits into two chiral primaries on reaching the boundary of 
the unitarity region. This implies that the massless fields made out 
of the $(h,Q)=(1/2,\pm 1)$ (anti-)chiral primary fields cannot acquire 
masses (= Liouville momenta) alone, without ``eating'' the fields 
from the $(h,Q)=(1,\pm 2)$ representations,\footnote{
This is a phenomenon reminiscent of the BPS saturation or the Higgs 
mechanism. We thank T.~Eguchi and S.-K. Yang for comments.
}
and both are trapped near the singularity. This will support the 
identification of the extra hypermultiplet as the massless soliton.
\vskip 1ex\noindent
\raisebox{-1mm}{\Huge $\cdot$} 
An $N=2$, $U(1)$ vector multiplet and a hypermultiplet
are also the massless excitations of the intersecting 
M5-branes\cite{HK}. This will agree with the duality of a conifold to the 
intersecting NS5-brane system proposed in refs.\cite{BVS,DM}.

\section{ALE Instantons: a Puzzle}
Let us next consider the ``blow-down'' limit of the ALE instantons. 
A similar guess from the defining equations suggests that the 
corresponding CFT will be a certain Landau-Ginzburg orbifold of 
a tensor product of $N\!=\!2$ minimal models and 
a $SL(2,\RR)/U(1)$ Kazama-Suzuki model with 
total central charge 6\cite{OV}. For example, the equations for the 
$A_{n-1}$-series are given by 
\begin{equation}
z_1^n+z_2^2+z_3^2=\mu z_4^{-n},
\end{equation}
where $z_i$ $(i=1,\ldots,4)$ are now the coordinates of 
$\PP^3$. Thus the corresponding CFT is the 
product of the minimal model of level $n-2$ and 
the $SL(2,\RR)/U(1)$ CFT of level $n+2$, 
{\it orbifolded by $\ZZ_n$}. 

We will now clarify what is the puzzle in the known formulas 
for the partition functions obtained in the previous works\cite{ES}.
(See also ref.\cite{Y}.)
The standard argument goes as follows: To construct a modular 
invariant, three different $N=2$ SCFTs must be taken into account: 
the free $N=2$ SCFT for the four transverse spacetime dimensions, 
the level-($n-2$) minimal model, and the level-($n+2$)
$SL(2,\RR)/U(1)$ SCFT. Since the 
$SL(2,\RR)/U(1)$ construction scans the whole 
$N=2$ unitarity region\cite{DLP}, we may consider any $U(1)$ charge 
lattice and sum over the representations in the third $c=3+\frac6n$ 
SCFT. Thus the building blocks must consist of three Jacobi theta 
functions (two from the transverse dimensions and one from  
$SL(2,\RR)/U(1)$), an $N=2$ minimal character,  
and a certain theta function. 

The crucial observation is that the following theta identity 
holds\cite{OV,ES} (See also ref.\cite{G}, eq.(2.24).): 
\begin{eqnarray}
&&\sum_{m\in\ZZsub_{\!2(k+2)}}
\!\!\Theta_{m,k+2}(\tau,\frac{kz}{k+2})
\left[\vartheta_3^3(0)
 ch^{(\mbox{\scriptsize NS})}_{l,m}(\tau,z)
 \phantom{\rule{0mm}{5mm}}
\right.\nonumber\\
&&~~~~~~~~~~~~~~~~~~~~~~~~~~~
\left.
-\vartheta_4^3(0)
 ch^{(\widetilde{\mbox{\scriptsize NS}})}_{l,m}(\tau,z)
-\vartheta_2^3(0)
 ch^{(\mbox{\scriptsize R})}_{l,m}(\tau,z)\right]\nonumber\\
&=&\chi_l^{(k)}(\tau,z)\left(
\vartheta_3^4(0)-\vartheta_4^4(0)-\vartheta_2^4(0)
\right), \label{NS5-K3}
\end{eqnarray}
where $ch_{l,m}(\tau,z)$ and $\chi_l^{(k)}(\tau,z)$ are the 
$N=2$ minimal and the affine $SU(2)$ characters, respectively. 
This means that if we take $\Theta_{m,k+2}(\tau,\frac{kz}{k+2})$ 
as the theta function for the $c=3+\frac6n$, 
$SL(2,\RR)/U(1)$ SCFT and consider the combinations 
of operators specified in the LHS of (\ref{NS5-K3}), 
we get a product of a level-$k$ $SU(2)$ WZW model 
and a free fermion theory. Modular invariants are easily constructed 
from those of the $SU(2)$ WZW models. This fact has been used to 
rationalize the duality between the singular K3 and the system of 
NS5-branes at the partition function level\cite{OV,ES}. 

Now here comes the puzzle: Since the lightest fields always come from 
the identity operator of $SU(2)$ while Jacobi's identity is 
independent of the type of the singularity, the ground-state 
degeneracy does not change whatever the singularity is! To look into 
the situation more closely, let us consider the simplest $A_1$ ($n=2$) 
case, the blow-down limit of the Eguchi-Hanson instanton. 
In this case the partition function is simply given by 
\begin{eqnarray}
&&Z_{\mbox{\scriptsize EH}}^{\mbox{\scriptsize (graviton)}}
=\int\frac{d^2\tau}{(\mbox{Im}\tau)^2}
\frac1{(\mbox{Im}\tau)^{\frac52}|\eta(\tau)|^{18}}
\cdot\left|
\frac12(\vartheta_3^4-\vartheta_4^4-\vartheta_2^4+\vartheta_1^4)
\right|^2\!\!.
\label{Z^graviton}
\end{eqnarray}
This is almost identical to the partition function for $D=10$ type~II(B) 
strings, but the essential difference is the number of $\eta$ 
functions; since the free $c=6$, $N=2$ SCFT for the four-dimensional 
flat space is replaced by the $SL(2,\RR)/U(1)$ SCFT, 
the partition function (\ref{Z^graviton}) has only {\it nine} 
(rather than twelve) 
$\eta$ functions in the denominator. Therefore, not the whole ghost 
ground-state energy $-1+\frac12=-\frac12$ is absorbed in the 
$\eta$ functions, but the $-\frac18$ due to the shortage of three 
$\eta$s is canceled by the ``Liouville ground-state energy'' 
of $+\frac18$ from the $SL(2,\RR)/U(1)$ internal 
SCFT. (Otherwise $q^{1/8}$ has no where to go and the modular 
invariance is lost. In the conifold case this was $q^{1/4}$ and 
can be seen as the gap above the origin (Figure 1).) Thus the graviton 
is again paired with a non-normalizable state of the internal SCFT, 
and hence acquires a $(\mbox{mass})^2$ of $\frac18$. This 
is always the case for other ADE singularities, being a common 
feature of non-critical strings formulated as such Gepner-like 
models.\footnote{
We thank M.~Bando, H.~Kawai and T.~Kugo for discussions 
on this point. 
} No other field is lighter than the graviton in this partition function. 
This implies that the modular invariant constructed from the identity 
(\ref{NS5-K3}) does {\it not} respect the geometry of the ALE spaces, 
{\it nor} can it see the zeromodes of the dual NS5-branes. This is the 
puzzle.

Is there any other orbit than (\ref{NS5-K3}) that closes under modular 
transformations? In the next section we will construct 
a new modular invariant partition function\footnote{The work done 
in collaboration with M.~Naka.} 
for the Eguchi-Hanson instanton by choosing a different initial 
condition in the $\beta$-method.  

\section{New Partition Function for the Eguchi-Hanson 
Instanton}
A general method to obtain a modular invariant combination from 
a set of theta functions has been given in Gepner's original 
paper\cite{G} and was called ``$\beta$-method''. We first note that 
the quartic polynomial of the theta functions in (\ref{Z^graviton}) 
can be written as follows: 
\begin{eqnarray}
&&~~(\vartheta_3+\vartheta_4)^2(\vartheta_3^2-\vartheta_4^2)
-(\vartheta_2+\vartheta_1)^2(\vartheta_2^2-\vartheta_1^2)\nonumber\\
&&+(\vartheta_3-\vartheta_4)^2(\vartheta_3^2-\vartheta_4^2)
-(\vartheta_2-\vartheta_1)^2(\vartheta_2^2-\vartheta_1^2)\nonumber\\
&&=2(\vartheta_3^4-\vartheta_4^4-\vartheta_2^4+\vartheta_1^4).
\end{eqnarray}
The LHS of the equation clearly shows how the GSO projection 
should be done in each sector; for each term, the first factor 
is the state sum 
of the transverse spacetime dimensions, while the second is the one 
coming from the internal $N=2$ SCFT. The first and the third terms 
are in the NS sector, and the second and the fourth ones are in the R
sector, for both spacetime and internal SCFTs. We now consider a variant 
of this equation: 
\begin{eqnarray}
&&~~(\vartheta_3+\vartheta_4)^2(\vartheta_2+\vartheta_1)^2
-(\vartheta_2+\vartheta_1)^2(\vartheta_3-\vartheta_4)^2\nonumber\\
&&+(\vartheta_3-\vartheta_4)^2(\vartheta_2-\vartheta_1)^2
-(\vartheta_2-\vartheta_1)^2(\vartheta_3+\vartheta_4)^2\nonumber\\
&&=16\vartheta_1\vartheta_2\vartheta_3\vartheta_4.
\end{eqnarray}
Each term gives the same phase in the modular $T$ transformation,
and their sum returns to itself up to a phase by the $S$ 
transformation. Using this, we write a new modular invariant:
\begin{eqnarray}
Z_{\mbox{\scriptsize EH}}^{\mbox{\scriptsize (doubleton)}}
&=&\!\int\!\frac{d^2\tau}{(\mbox{Im}\tau)^2}\!
\frac1{(\mbox{Im}\tau)^{\frac52}|\eta(\tau)|^{18}}
\cdot\!\frac1{16^2}\nonumber\\
&&\cdot\mbox{\LARGE $|$}(\vartheta_3+\vartheta_4)^2
\left((\vartheta_2+\vartheta_1)^2
 +(\vartheta_2-\vartheta_1)^2\right)\nonumber\\
&&
-(\vartheta_2+\vartheta_1)^2
\left((\vartheta_3-\vartheta_4)^2
 +(\vartheta_3+\vartheta_4)^2\right)\label{Z^doubleton}\\
&&
+(\vartheta_3-\vartheta_4)^2
\left((\vartheta_2-\vartheta_1)^2
 +(\vartheta_2+\vartheta_1)^2\right)\nonumber
\\
&&\raisebox{3mm}{$-(\vartheta_2-\vartheta_1)^2
\left((\vartheta_3+\vartheta_4)^2
 +(\vartheta_3-\vartheta_4)^2\right)$}
\raisebox{2mm}{\LARGE $|$}^{\raisebox{-1mm}{\scriptsize $2$}}\!\!\!.
\nonumber
\end{eqnarray}
In fact, the alternating sum inside $|\cdots |$ 
vanishes {\it trivially}; however, it can be interpreted as a 
consequence of the cancellation between the NS- and the 
R-sectors. Again, the first and the third terms are in the NS-sector
for the transverse spacetime SCFT, but they are now paired with the 
R-sector of the internal SCFT (and similarly for the transverse 
R-sector)!\footnote{The change of the fermion boundary condition was 
discussed in ref.\cite{ABFZG}, but they were led to a 
different result.} Note that 
$Z_{\mbox{\scriptsize EH}}^{\mbox{\scriptsize (graviton)}}$ 
and 
$Z_{\mbox{\scriptsize EH}}^{\mbox{\scriptsize (doubleton)}}$ 
are separately modular invariant.

Remarkably, 
$Z_{\mbox{\scriptsize EH}}^{\mbox{\scriptsize (doubleton)}}$ does 
not contain any graviton because of its peculiar GSO projection.
The lightest IIB fields are combined into a $(2,0)$ tensor 
multiplet in six dimensions, which coincides with the zeromode 
excitations on the IIA NS5-brane\cite{CHS}!; this is a manifestation 
of T-duality.
In fact, its $(\mbox{mass})^2$ is {\it negative} ($=-\frac18$)!  
It does not necessarily mean the instability of our vacuum because 
there is no reason to believe that the six-dimensional spacetime 
is flat any more. Perhaps it might be understood as a 
doubleton-like mode. The doubleton\cite{doubleton} is known to be 
the lowest Kaluza-Klein mode with a negative $(\mbox{mass})^2$ 
in (say) the $AdS_7\times S^4$ compactification of $D=11$ supergravity. 
It is a pure gauge mode in the bulk but has a holographic dual 
on the six-dimensional $AdS$ boundary, on which an M5-brane sits. 
If the T-duality relation between a singular ALE and 
NS5-branes persists even in strong coupling, and if the abstract 
CFT approach can consistently describe physics near the singularity 
 (as it seemed to be in the conifold case), the CFT must ``see'' 
some alternative dual to the strongly coupled type~IIA theory; 
the one-loop CFT calculation for the latter itself is 
certainly inconsistent. Thus if the NS5-branes could be replaced 
by M5-branes, the mysterious tachyonic fields would then have a  
explanation as a doubleton-like mode. For the full understanding 
of this, we will need to generalize our formula to other ADE cases.

\section{Concluding Remarks}
We have constructed partition functions for type II strings on 
a conifold and a singular Eguchi-Hanson instanton by using a 
Gepner-like abstract CFT approach. 

In the conifold case, we found 
an extra massless hypermultiplet and argued that this might be 
identified as the massless soliton. The appearance of the massless 
soliton in the spectrum comes as a surprise; to confirm the 
identification further, the nature of the extra states must be 
clarified in the boundary-state formulation of D-branes.

The doubleton-like mode in the Eguchi-Hanson case is also quite 
unexpected. The relation to the M-theory dual is still mysterious. 
It would be interesting to compare (reconcile?) our result with 
the works of refs.\cite{GK}.

\section*{Acknowledgments}
We thank M.~Naka and T.~Tani for valuable discussions.


\begin{thebibliography}{99}
\bibitem{S}
A.~Strominger,
{\it Nucl. Phys.} {\bf B451}, 96 (1995).

\bibitem{W}
E.~Witten,
{\it Nucl. Phys.} {\bf B443}, 85 (1995).

\bibitem{BVS}
M.~Bershadsky, C.~Vafa and V.~Sadov, 
{\it Nucl. Phys.} {\bf B463}, 398 (1996). 

\bibitem{M}
S.~Mizoguchi, 
{\it JHEP} {\bf 0004}, 014 (2000). 

\bibitem{OV}
H.~Ooguri and C.~Vafa,
{\it Nucl. Phys.}  {\bf B463}, 55 (1996). 

\bibitem{ES}
T.~Eguchi and Y.~Sugawara, 
{\it Nucl. Phys.} {\bf B577}, 3 (2000). 

\bibitem{GV}D.~Ghoshal and C.~Vafa,
{\it Nucl. Phys.}  {\bf B453}, 121 (1995).

\bibitem{DLP}
L.J.~Dixon, M.E.~Peskin and J.~Lykken,
{\it Nucl. Phys.} {\bf B325}, 329 (1989).


\bibitem{N=2Liouville}
O.~Aharony, M.~Berkooz, D.~Kutasov and N.~Seiberg,
{\it JHEP} {\bf 9810}, 004 (1998). \\
A.~Giveon, D.~Kutasov and O.~Pelc,
{\it JHEP} {\bf 9910}, 035 (1999).

\bibitem{GK}A.~Giveon and D.~Kutasov,
{\it JHEP} {\bf 9910}, 034 (1999); 
{\it JHEP} {\bf 0001}, 023 (2000). 

\bibitem{BG}
A.~Bilal and J.L.~Gervais,
{\it Nucl. Phys.} {\bf B284}, 397 (1987).

\bibitem{ET}
T.~Eguchi and A.~Taormina,
{\it Phys. Lett.} {\bf B210}, 125 (1988).

\bibitem{HK}A.~Hanany and I.R.~Klebanov,
{\it Nucl. Phys.} {\bf B482}, 105 (1996).

\bibitem{DM}
K.~Dasgupta and S.~Mukhi,
{\it Nucl. Phys.}  {\bf B551}, 204 (1999).

\bibitem{Y}S.~Yamaguchi, hep-th/0007069.

\bibitem{G}D.~Gepner,
{\it Nucl. Phys.} {\bf B296}, 757 (1988).

\bibitem{ABFZG}
D.~Anselmi, M.~Billo, P.~Fre, L.~Girardello and A.~Zaffaroni,
{\it Int. J. Mod. Phys.} {\bf A9}, 3007 (1994).

\bibitem{CHS}
C.G.~Callan, J.A.~Harvey and A.~Strominger,
{\it Nucl. Phys.} {\bf B367}, 60 (1991).

\bibitem{doubleton}
M.~Gunaydin, P.~van Nieuwenhuizen and N.P.~Warner,
{\it Nucl. Phys.} {\bf B255}, 63 (1985).


\end{thebibliography}
\end{document}